# Fault-Tolerant Directional Couplers for State Manipulation in Silicon Photonic-Integrated Circuits

*Moshe Katzman[1†], Yonatan Piasetzky[2†], Evyatar Rubin[1], Ben Birenboim[1], Maayan Priel[1], Avi Zadok[1], and Haim Suchowski[2]*

† The first two authors contributed equally to this work

[1]*Faculty of Engineering and Institute for Nano-Technology and Advanced Materials, Bar-Ilan University, Ramat-Gan 5290002, Israel*

[2]*Department of Condensed Matter Physics, School of Physics and Astronomy, Tel Aviv University, Tel Aviv 6997801, Israel*

**Abstract**:

Photonic integrated circuits play a central role in current and future applications such as communications, sensing, ranging, and information processing. Photonic quantum computing will also likely require an integrated optics architecture for improved stability, scalability, and performance. Fault-tolerant quantum computing mandates very accurate and robust quantum gates. In this work, we demonstrate high-fidelity directional couplers for single-qubit gates in photonic integrated waveguides, utilizing a novel scheme of detuning-modulated composite segments. Specific designs for reduced sensitivity to wavelength variations and real-world geometrical fabrication errors in waveguides width and depth are presented. Enhanced wavelength tolerance is demonstrated experimentally. The concept shows great promise for scaling high fidelity gates as part of integrated quantum optics architectures.

## 1. Introduction

Photonic-integrated circuits (PICs) are essential for present and future data center communications [1]. PICs are also widely regarded among the most promising platforms for the realization of quantum technologies in computation, information processing, and sensing [2–5]. The advancement of quantum integrated photonics would critically depend on the reduction of errors [2,4]. Compared with a classical application, quantum technologies are far less tolerant to possible uncertainties associated with the fabrication of PICs [6].

The directional coupler is among the most fundamental and widely employed building blocks of PICs [5]. Quantum photonic applications require that the splitting ratios of directional couplers comply with target design to the fourth decimal point [6]. This requirement places stringent fabrication tolerances on process parameters such as etching depth, waveguides widths, etc., which are difficult to meet in practice. Geometrical errors are often correlated: they affect both constituent waveguides in the directional coupler in the same manner. In addition, quantum-photonic integrated circuits would need to operate over ranges of wavelengths and temperatures, leading to further variations in their transfer functions. Robust directional couplers are, therefore, a prerequisite for quantum-photonic integrated circuits. Unfortunately, standard directional couplers with uniform waveguide sections do not exhibit sufficient tolerance to inevitable fabrication and spectral uncertainties [5].

Much effort has been dedicated in recent years to the design and realization of robust photonic-integrated couplers and quantum gates [2,7–9]. For example, several works have reported wavelength insensitive couplers in silicon-on-insulator (SOI) photonics [10–13]. However, fewer works addressed tolerance to geometrical errors in fabrication. Recently, some of us have proposed a design concept for fault-tolerant directional couplers based on waveguide sections of different widths [14]. The approach is based on the principles of composite pulse sequences used in atomic physics [15,16]. The so-called composite-sections couplers are simple to realize, and they do not require modifications to the fabrication protocols. The design approach is generic and can be optimized for the mitigation of errors in wavelength or geometry.

In this work, we report composite-sections directional couplers in the standard SOI platform, the workhorse of PICs. One specific design addresses wavelength variations, whereas another provides enhanced tolerance to real-world geometrical fabrication errors in the widths and etching depth of both coupler waveguides. Improved spectral stability is demonstrated experimentally: the coupling ratios of hundreds of microns-long uniform and composite-sections couplers are characterized across the C-band wavelengths. The spectral sensitivity of the composite-sections couplers splitting ratio is reduced by over an order of magnitude. The composite-sections couplers may serve as a fundamental building block of practical quantum silicon photonics.

## 2. Design of composite-section directional couplers

According to the coupled-mode theory, the propagation of the pair of electrical fields $E_{1,2}$ in a directional coupler of a fixed cross-section is described by the following unitary propagator:

$$\begin{bmatrix} E_1(z) \\ E_2(z) \end{bmatrix} = \begin{bmatrix} \cos\left(\frac{A}{2}\right) + \frac{i\Delta\beta}{\kappa_g}\sin\left(\frac{A}{2}\right) & -\frac{i\kappa}{\kappa_g}\sin\left(\frac{A}{2}\right) \\ -\frac{i\kappa}{\kappa_g}\sin\left(\frac{A}{2}\right) & \cos\left(\frac{A}{2}\right) - \frac{i\Delta\beta}{\kappa_g}\sin\left(\frac{A}{2}\right) \end{bmatrix} \begin{bmatrix} E_1(z_0) \\ E_2(z_0) \end{bmatrix} = \widehat{U}\begin{bmatrix} E_1(z_0) \\ E_2(z_0) \end{bmatrix} \quad (1)$$

Here κ is the coupling coefficient between the two waveguides, determined mainly by the distance between the cores. $\Delta\beta = \frac{\beta_1 - \beta_2}{2}$ denotes the detuning, or the phase mismatch parameter, which is determined by the propagation constants $\beta_{1,2}$ in the two constituent waveguides. Also, in Eq. (1), $z$ represents the axial coordinate in the direction of propagation, $z_0$ is the point of entry into the coupler, $\kappa_g = \sqrt{\kappa^2 + \Delta\beta^2}$ is the generalized coupling coefficient, and $A = \int_{z_0}^{z} \kappa_g \, dz = \kappa_g(z - z_0)$ is the integral of the generalized coupling coefficient, also referred to as the 'pulse area'.

In the design of broadband, robust composite-section directional couplers, several segments of uniform waveguides widths and uniform gap are concatenated together. Each is described by its own unitary matrix, such as to achieve better robustness for a specific error model. We relate to variations in the operating wavelength first. To realize couplers with reduced spectral sensitivity, we use two sections, $i = 1,2$, with $\kappa_1 = \kappa_2 = \kappa$, $\Delta\beta_1 = -\Delta\beta_2 = \kappa$. We utilize the first-order scheme [14], in which the second-order derivative with respect to pulse area error is nullified, as defined in equation **Error! Reference source not found.**:

$$\frac{\partial^2}{\partial A^2}\left(\prod_i \widehat{U}_i\right) = 0 \quad (2)$$

A two-section composite directional coupler is illustrated in Fig. 1. The coupler comprises one uniform waveguide and a second waveguide of position-dependent width. The second waveguide consists of repeating sections of widths $W_{1,2}$ and length $L_{1,2}$. The lateral separation between the centers of the two waveguides remains constant. To reduce reflection losses at abrupt width changes, a 1 μm long tapered waveguide is inserted in each width transition.

Figure 2 shows the calculated power splitting ratios of directional couplers between two SOI ridge waveguides, designed nominally for complete power swap, as functions of wavelength. The thickness of the silicon device layer is 220 nm and that of the buried oxide layer is 2 μm. The exact dimensions of waveguides within the couplers were taken from atomic force microscope measurements of fabricated devices (see next section). The depth of partial etching is 77 nm between the two ridge waveguides and 85 nm to their sides. The lateral separation between the centers of the two waveguides is fixed at 1 μm. The propagation constants $β_{1,2}$, coupling coefficient κ and mismatch parameter Δβ were calculated numerically for each waveguide segment, and the coupling coefficients of the directional couplers were found through applying Eq. (1) to each section. The dashed red trace presents a 32.9 μm-long uniform coupler between waveguides of 750 nm widths. Corresponding results for a composite-sections coupler ($W_1 = 610$ nm, $W_2 = 835$ nm, $L_1 = 24.4$ μm, $L_2 = 21.8$ μm, overall length of 46.22 μm) are shown in solid blue. The composite-sections design reduces wavelength dependent errors in the coupling ratio from 0.25% to 0.1%.

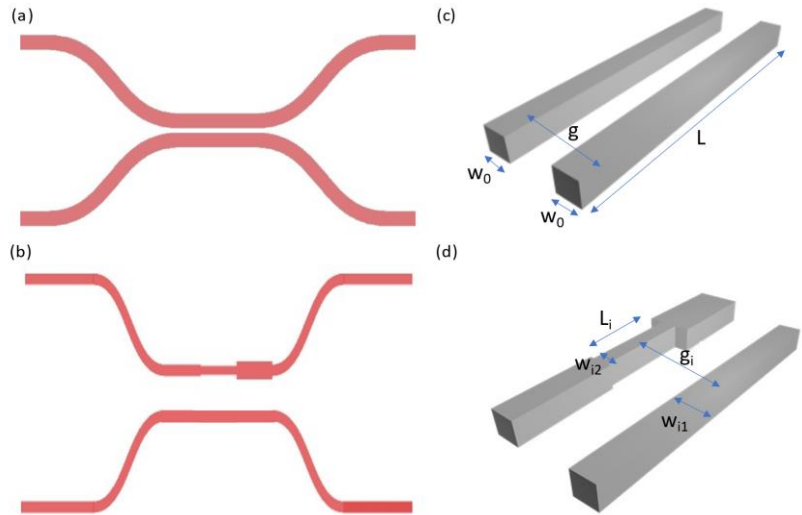

Figure 1. Schematic illustration of standard and composite-sequence directional couplers. (a): Top view of a standard coupler with constant cross-section vs. composite-sequence coupler (b) which has piecewise constant cross-section for any number of sequences. (c-d) Shows a 3D view of the coupling region inside the standard and composite couplers respectively. The design parameters of waveguides widths, lengths, and gaps in the different segments are shown (see text).

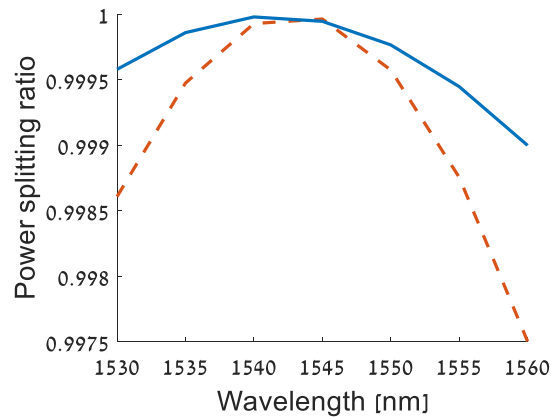

Figure 2. Calculated power splitting ratios of silicon-photonic directional couplers as functions of wavelength. Dashed red: Standard coupler comprised of two uniform ridge waveguides with 750 nm widths and 1 μm separation between their centers. The ridges are defined in a 220 nm-thick device layer through partial etching to a depth of 77 nm between the waveguides and 85 nm to their sides. The coupler is 32.9 μm long. Solid blue: A composite-sections coupler of dissimilar ridge waveguides with the same etching depths. One waveguide is uniform, with a 750 nm width. The width of the second waveguides varies between two sections of 24.39 μm and 21.83 μm lengths. The widths in the first and second sections in each period are 660 nm and 890 nm, respectively. The separation between the centers of the two waveguides is fixed at 1 μm. The calculations suggest that composite-sections coupler design is more robust to wavelength variations.

Next, we propose a composite-sections coupler design with increased tolerance to real-world geometric fabrication errors. The most commonly encountered deviations from design parameters are in the partial etching depth of ridge waveguides and in their widths. Both types of deviations are likely to be the same for the two constituent waveguides of a directional coupler, along its entire length. Errors in width and depth are statistically uncorrelated. Referring to Eq. (1), our analysis suggests that the detuning parameter $\Delta\beta$ is the most sensitive to geometrical errors. A composite-section coupler was designed to cancel out the detuning error to the first order.

Figure 3 shows the calculated splitting ratios of two directional couplers, designed again for a complete power swap, as functions of width and etching depth errors. Panel (a) presents a standard coupler of uniform cross-section, whereas panel (b) shows the results for a three-section composite coupler (for details of sections lengths and widths, see Fig. 3). Etching depth errors $\delta h$ between ±4 nm and width errors $\delta w$ within ±40 nm are considered. The analysis suggests that the composite-sections coupler is more robust to geometrical fabrication errors. Compared with the uniform coupler, the splitting ratio remains above 99.7% over $\delta h$ and $\delta w$ values that are three times larger (see contour lines in Fig. 3).

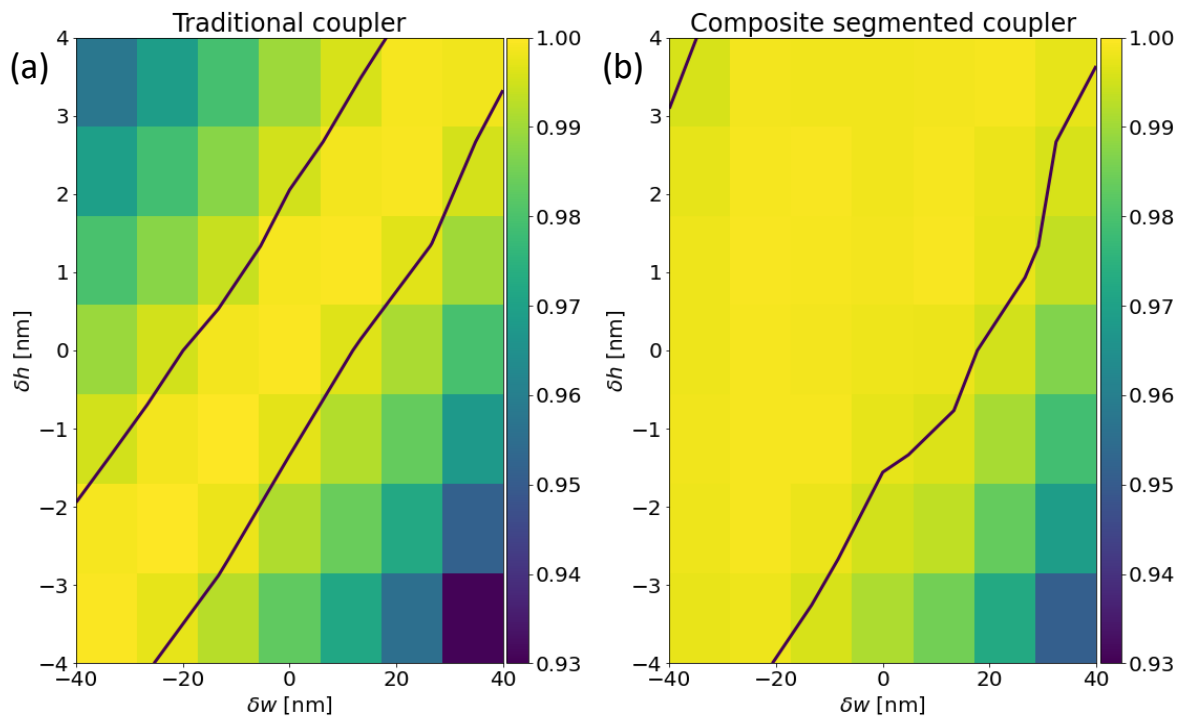

*Figure 3. Calculated power splitting ratios of silicon-photonic directional couplers as functions of errors $\delta h$ in the partial etching depth and errors $\delta w$ in the widths of constituent ridge waveguide. The ridges are defined in a 220 nm-thick device layer through partial etching to a nominal depth of 70 nm. The black lines mark a contour of splitting ratios 99.7% and higher. The errors $\delta h$ and $\delta w$ are not correlated, and both are common to the two waveguides throughout the coupler length. (a): Standard coupler comprised of two uniform ridge waveguides with a nominal width of 1200 nm and separation of 1550 nm between the waveguide centers. The coupler is 157 μm long. (b): A composite-sections coupler of dissimilar ridge waveguides with the same etching depth. One waveguide is uniform, with a nominal 1200 nm width. The designed width of the second waveguides varies in three composite segments: 1137 nm width for 119 μm, 1200 nm width over 157 μm, and 1285 nm width for 101 μm. The separation between the centers of the two waveguides is again fixed at 1550 nm. The calculations suggest that the composite-sections coupler is more robust to real-world geometrical fabrication errors.*

## 3. Device fabrication and testing

To compare between uniform and composite-section couplers, test patterns of racetrack resonators were fabricated in standard SOI substrates. The lengths of the resonators were 1.3 mm. The widths of waveguides within the directional couplers forming the resonators, the separation between the waveguides, and the etching depths were the same as those in Fig. 2 above. A reference resonator device was based on a uniform coupler of 350 μm length. A second device included a composite-sections coupler with ten repeating periods. Following the ten periods, another composite section was added to provide critical coupling into the resonator. That section comprised of two segments, with the width of one waveguide fixed at 750 nm. The width of the other waveguide was 486 nm in the first section and 690 nm in the second. The lengths of the two sections were 11.6 μm and 32.86 μm, respectively. Long couplers were deliberately used in both the uniform and composite section designs, to enhance the wavelength dependence of their splitting ratios and compare their robustness. Errors in the long couplers are also indicative of the anticipated performance trends of multiple cascaded gates, in more complex circuits.

Optical waveguides were defined in the silicon device layer using electron-beam lithography, followed by inductively coupled plasma reactive-ion etching. The etching process used a mixture of $SF_6$ and $C_4F_8$ gasses, at flow rates of 65 sccm and 10 sccm, respectively. Etching was carried out at a vacuum level of $4 \times 10^{-10}$ bar and radio frequency power of 100 W at a 6 nm×s$^{-1}$ rate. Figure 4(a) shows a top-view optical microscope image of one resonator device. Figure 4(b) presents a top view optical microscope image of one period within the composite-sections coupler. Figures 4(c)-4(e) show scanning electron microscope images of focused ion beam cross-sections of uniform and composite-section couplers. Light was coupled between the bus waveguides of resonator devices and standard single-mode fibers using vertical grating couplers. The coupling losses were 10 dB per facet.

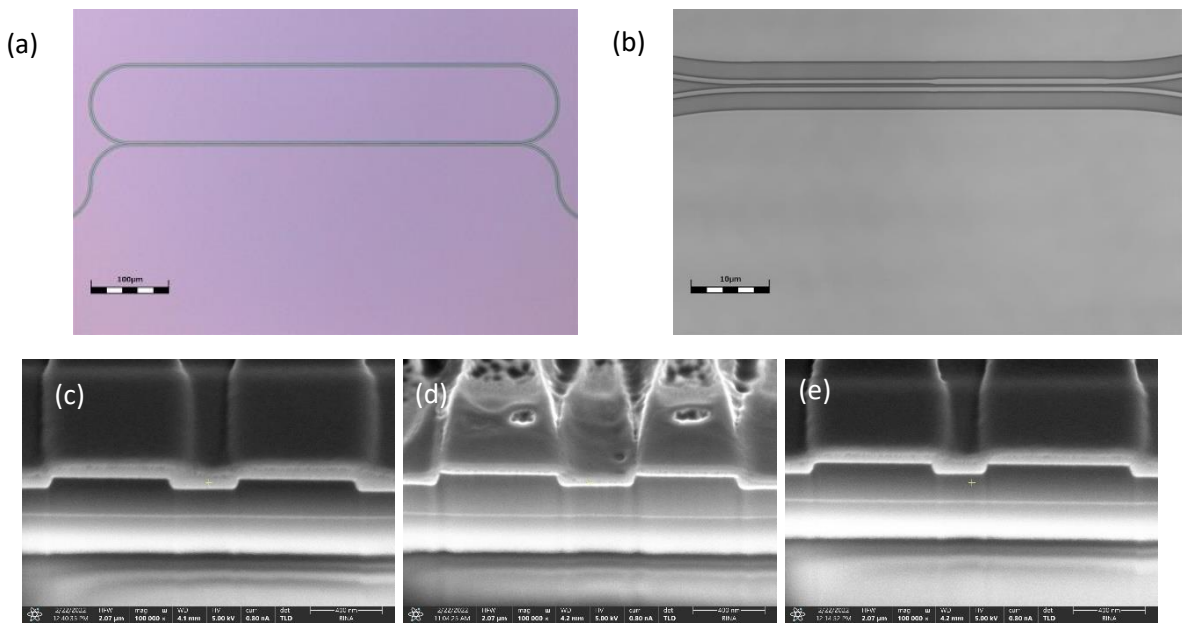

Figure 4. Fabricated devices (a): Top-view optical microscope image of a silicon-photonic racetrack resonator device used in the comparison between uniform and composite-sections directional couplers. Scale bar represents 100 μm. (b): Top-view optical microscope image of one period within a long composite-sections directional coupler. Scale bar represents 10 μm. (c)-(e): Scanning electron microscope images of focused ion beam cross-sections of directional couplers. The scale bars represent 400 nm. (c): Uniform coupler with two identical waveguides, each 750 nm wide. (d): A first section of a composite coupler, in which the width of the right waveguide is reduced to 610 nm. (e): A second section of a composite coupler, in which the width of the right waveguide is increased to 835 nm.

Figure 5(a) shows optical vector network analyzer measurements of the transfer functions of optical power through the two devices. The wavelength resolution was 3 picometers. The transfer functions consist of multiple periods with a free spectral range of 0.5 nm. The Q factor of the resonators is 30,000. The extinction ratios of the periodic transmission notches of the uniform-coupler device vary strongly across the C-band wavelengths range, between 0.25 dB and 20 dB. By contrast, the extinction ratios of the resonator with the composite-sections coupler remained within $16 \pm 3$ dB over the entire wavelength range.

Figure 5(b) presents the power splitting ratios of the two couplers as functions of wavelengths, as calculated from the observed extinction ratios of the two transfer functions [5]. The splitting ratio of the long, uniform coupler varies between 0.3 and 1, whereas that of the composite-sections coupler remains within 0.34 ± 0.02. The results illustrate the superior robustness of the composite-sections design with respect to wavelength changes, in agreement with analysis [14]. Fig. 5(b) also presents the calculated splitting ratios of both directional couplers as functions of wavelengths. Contributions of transition regions leading the two waveguides to and from the coupling region were accounted for as well. Agreement between models and measurements of splitting ratios is very good.

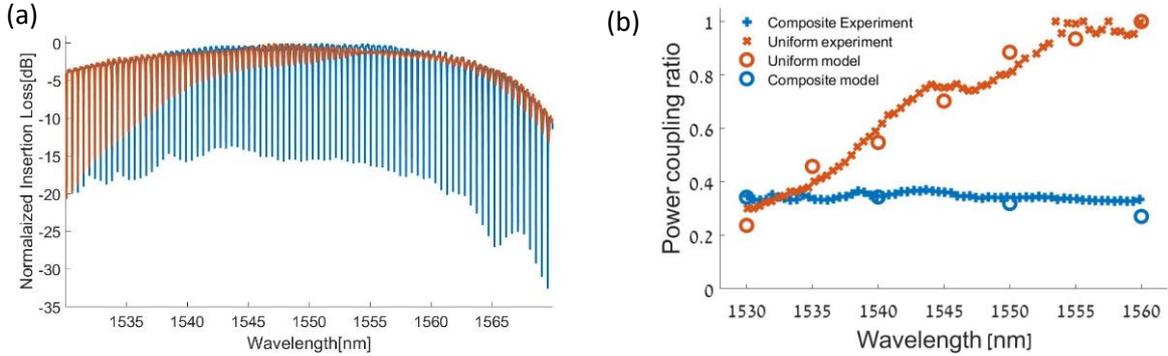

Figure 5. (a): Measured transfer functions of optical power through two racetrack resonator devices in silicon-on-insulator, as functions of wavelength. Red: A reference device comprised of a long uniform directional coupler. Blue: Device based on a long composite-sections directional coupler. The reference device exhibits strong variations in extinction ratios of periodic transmission notches across the C-band wavelengths. The extinction ratios of the composite-sections coupler device remain more uniform across the same wavelength range. (b): Splitting ratios of the two-directional couplers, based on the measured extinction ratios of panel (a) (same colors, see legend). The composite-sections design reduces the wavelength variations of the coupler splitting ratio by a factor of 15. The splitting ratios agree well with calculations based on Eq. (1).

## 4. Summary

In this work, we have proposed and demonstrated fault-tolerant directional couplers comprised of sections with different waveguides widths. The design concept is inspired by the principles of composite pulses in atomic physics [15,16], which were recently expanded by the detuning modulated composite sequence scheme to tackle errors in PICs [14]. This study shows that the scheme developed for a general quantum system is also applicable to integrated optics. The results signify considerable improvement in the robustness of single-qubit state manipulation in the common path encoding of photonic qubits.

We have experimentally demonstrated the enhanced wavelength insensitivity of a full swap coupler. Compared with a reference uniform device, splitting ratio variations of a long composite-sections coupler across the C-band were reduced fifteen-fold. Measurements and calculations were in good agreement. The results suggest large potential benefit in cascaded operation of multiple couplers in series within larger processing layouts. Analysis showed that wavelength variations in a short full-swap coupler are reduced from 0.25% in the uniform cross-section device to 0.1% in the two-sections layout. A second design targeted common errors in both waveguides' width and etching depth within the coupler. The width and depth errors for which the splitting ratio remain above 99.7% were calculated for a reference uniform coupler and a composite coupler. The composite coupler design tolerates geometrical errors that are three times larger.

The design concept shows great promise for enabling and scaling integrated photonic quantum information processors, which will require many single-qubit operations in every architecture. While this design can eliminate errors to the first order, the concept can be extended to mitigate second- and higher-order errors as well. Additional extensions of this scheme would address the design of robust single- and two-qubit unitary gates.